\documentclass[12pt]{iopart}

\usepackage{amsmath}
\usepackage{graphicx}
\usepackage{amssymb}
\usepackage{cite}

\begin{document}


\title[]{Efficient post-processing of electromagnetic plane wave simulations to model arbitrary structured beams incident on axisymmetric structures}

\author{J. J. Kingsley-Smith and F. J. Rodr\'iguez-Fortu\~no}
\address{Department of Physics and London Centre for Nanotechnology, King's College London, Strand, London, WC2R 2LS, UK}
\ead{francisco.rodriguez\_fortuno@kcl.ac.uk}
\vspace{10pt}
\begin{indented}
\item[]February 2023
\end{indented}

\begin{abstract}

\noindent The study of an optical beam interacting with material structures is a fundamental of nanophotonics. Computational electromagnetic solvers facilitate the rapid calculation of the scattering from material structures with arbitrary geometry and complexity, but have limited efficiency when employing structured excitation fields. We have developed a post-processing method and package that can efficiently calculate the full three-dimensional electric and magnetic fields for any optical beam incident on a particle or structure with at least one axis of continuous rotational symmetry, called an axisymmetric body (such as a sphere, cylinder, cone, torus or surface). Provided an initial batch of plane wave simulations is computed, this open-source package combines data from computational electromagnetic solvers in a post-processing fashion using the angular spectrum representation to create arbitrarily structured beams, including vector vortex beams. Any and all possible incident beams can be generated from the initial batch of plane wave simulations, without the need for further simulations. This allows for efficiently performing parameter sweeps such as changing the angle of illumination or translating the particle position relative to the beam, all in post-processing, with no need for additional time-consuming simulations. We demonstrate some applications by numerically calculating optical force and torque maps for a spherical plasmonic nanoparticle in a tightly focused Gaussian beam, a plasmonic nanocone in an azimuthally polarised beam and compute the fields of a non-paraxial Laguerre-Gaussian vortex beam reflecting on a multilayered surface. We believe this package, called BEAMS, is a valuable tool for rapidly quantifying electromagnetic systems that are beyond traditional analytical methods. 

\end{abstract}

%
\vspace{2pc}
\noindent{\it Keywords}: beam physics, nanoparticle, scattering, surface, symmetry, optical force, optical torque
%
%
\maketitle
%
%

\section{Introduction}

Ever since the invention of the laser in the mid-twentieth century, there has been immense progress in the field of wavefront engineering. The precise design and manipulation of optical beams has found applications in most areas of science, including physics, biology, chemistry, telecommunications, quantum computing and medicine \cite{Abbott2016,Moffitt2008,Righini2009,Brown2011,Svoboda1994,Kuo2001,Juan2011,Greulich2017,Xin2019,Bustamante2020,Grier2003,Ashkin2000,Marago2013,Ren2021,Guo2013,Babiker2019}. Collimated electromagnetic fields are rich with potential because of the variety of ways in which they can be sculpted and generated. The field intensity can manifest in the form of high order Gaussian modes like the Laguerre-Gaussian and Hermite-Gaussian modes, and these alone can incite complex phenomena such as orbital angular momentum \cite{Allen1999,Yao2011,Curtis2003,Padgett2011,Shen2019}. Beyond intensity variations, the polarisation state of the beam can be made inhomogeneous to create what is known as a vectorial vortex beam \cite{Zhan2009,Hanifeh2020}. These beams are now readily reproducible in laboratory conditions \cite{Wang2018b,Wang2019,Zhu2019,Tidwell1993,Fu2016,Cardano2012,He2022,Mandal2020,Lerman2010,Kasperczyk2015} and present exciting new ways in which matter can interact with a light field. Their structure have already found many novel applications including particle orientation analysis, magnetic field localisation and inducing orbital motion in an isotropic particle \cite{Novotny2001,Kasperczyk2015,Wang2010}. 

It is only natural to combine sophisticated vortex beams with complex material structures in the pursuit of unknown effects and future technologies. Modern nanofabrication techniques have enabled the realisation of a rich variety of nanostructures, each of which have properties that can be applied to drug delivery, sensing, nanochemistry and photocatalyis \cite{Singh2009,McPhillips2010,DeAngelis2013,Wong2013,Sershen2000,Wang2019a,Kale2014}. The study of these structures and their interactions with light is crucial for optimisation and further development. Whilst analytical methods such as Mie theory \cite{Bohren1998,VanderHulst2003} and Green's method for multipoles \cite{Sipe1987,Novotny2006,Vazquez-Lozano2019} do exist for calculating simple cases, many structures are not accurately portrayed by these methods and often require numerical finite element solutions. This is particularly relevant when a strongly interacting substrate is present, which is commonplace in experimental measurements. However, many commercial solvers either do not support, or require elaborate steps and big simulation regions, to simulate sophisticated optical illumination fields. By contrast, plane wave simulations are always fast, efficient and simple. This limitation usually dramatically slows down the study of tight-focusing regimes and the fascinating physics that lies therein. 

In this paper, we demonstrate a numerical approach to solving many systems often encountered in nanophotonics whilst accommodating complete freedom over the illumination properties. The advantage of this approach is that complex beams are generated as a post-processing step purely from plane wave simulations. We combine the generality of a finite element solver with the power of symmetry to generate full-field 3D solutions that can then be used to calculate any optical quantities desired, such as optical forces, optical torques, power flow, spin density, helicity and momentum flow. The only constraint is that the system must be axisymmetric; the matter's geometry must possess one axis of continuous rotational symmetry. Otherwise, the method is still valid but unlikely to be efficient. Demonstrations involving a core-shell particle, a plasmonic nanocone and a multilayer reflection problem are analysed with our method as examples to illustrate its generality, each illuminated with focused optical beams of varying structures. 

\section{Methods}

This section begins by introducing the angular spectrum approach and its formalism. We explain how to represent any beam with this approach and then proceed to directly incorporate the scattering of a spherical particle. This is then extended to a scattering axisymmetric body and concludes with some expressions for optical force and optical torque that are used in the Results section. 

\subsection{Adding structure scattering to an angular spectrum}\label{subsec:addstructure}

The first principle behind our method lies in the angular spectrum approach for an electromagnetic field of angular frequency $\omega$ which can be expressed as,
\begin{equation}\label{eq:angularspectrum}
    \mathbf{E}(\mathbf{r}) = \int^{2\pi}_{\phi=0}\int^{\pi}_{\theta=0} \mathcal{A}(\mathbf{k}) \, e^{i \, \mathbf{k} \cdot (\mathbf{r}-\mathbf{r}_0)} \, \underbrace{\text{sin}(\theta) \, \text{d}\theta \, \text{d}\phi}_{\text{d}\Omega_k}, 
\end{equation}
where $\mathbf{E}$ is the electric field at a position $\mathbf{r} = (x,y,z)$, $\mathbf{r}_0$ determines the phase centre of the optical field\footnote{When the optical field is a beam, $\mathbf{r}_0$ is the focus of the beam.}, $\mathcal{A}$ is the electric field's angular spectrum which is dependent on the wavevector $\mathbf{k} = k\big(\text{sin}(\theta) \, \text{cos}(\phi),\text{sin}(\theta) \, \text{sin}(\phi), \text{cos}(\theta)\big)$, and $\Omega_k$ is the solid angle of the k-sphere defined by all possible orientations of $\mathbf{k}$.

The angular representation indicates that any monochromatic electromagnetic field distribution can be decomposed into a linear combination of plane waves with varying spatial frequencies. One can be sure that any linear combination of plane waves will create another exact solution of Maxwell's equations and vice versa, any exact solution of Maxwell's equations (in linear materials) can be produced as a superposition of plane waves. 
We can further conceptualise the angular spectrum of any optical beam in free-space as a collection of plane waves that can all be mapped onto a single common propagating plane wave via some 3D rotations. This mapping is augmented by the angular spectrum in order to change the phase and magnitude of each plane wave as they are rotated. Instead of decomposing a known beam's field distribution, one can also conduct the inverse procedure, starting from a single plane wave, copy it a large number of times and both rotate and modulate the various copies according to the previously mentioned mapping. All the plane waves can be integrated and the resultant linear combination is equal to the desired optical beam. 

If a spherical particle is added to the beam system, the fields are perturbed by the scattering. However, if we again take a single plane wave, this time incident on the spherical particle, and rotate the total vector field (complete with the sphere's scattering), we know that the solution is physically identical to the unrotated version because the particle is rotationally symmetric. Therefore with the simulation of a single plane wave incident on the spherical particle, we can repeat the process of rotating the fields and mapping the results to an angular spectrum of a beam, transforming this plane wave scattering system into the field distribution of a beam incident on the spherical particle. From this, we gain a strikingly general conclusion. Given a single simulation of a spherical particle with an incident plane wave, we can transform the plane wave into any optical illumination that we desire in simple post-processing steps. The rotational symmetry of the system reduces the number of required plane wave simulations (PWS) from infinite to just one. This is the second principle behind our method. Whilst this alone can still prove useful for simulating many common photonic systems, we can further extend this principle to cylinders, cones, tori, and any other particle shapes that possess continuous rotational symmetry (i.e. any axisymmetric body\cite{Peterson1997}) as follows. 

We choose the convention that the particle is orientated such that an axis of rotational symmetry is along the $z$-axis ($\theta = 0$) and a plane wave is incident at an azimuthal angle $\phi$ and an angle from the $z$-axis $\theta$. The entire system can be rotated around the $z$-axis (i.e. varying the azimuthal angle $\phi$) and still remain physically identical. However, any change in $\theta$ will not follow any symmetry in the system and so the scattering will be different. This indicates that a new simulation is needed for every value of $\theta$, whilst all plane waves whom share the same value of $\theta$ are related to each other by a simple rotation about $\phi$. 
We therefore need to sample a number of simulations with different values of $\theta$ to accurately extract the change in the scattering behaviour as $\theta$ changes. When compared with the general case where the particle has no symmetry at all and a large number of PWS would be required, we can still reduce the number of simulations needed significantly because the rotational symmetry effectively reduces the dimensionality of the problem.

Before we can apply symmetry arguments to simplify the problem, we must first be able to place particle scattering into a beam's fields via the angular spectrum approach. We can mathematically apply this to Eq.~(\ref{eq:angularspectrum}) by first expanding $\mathcal{A}$ into an orthogonal basis. We use a spherical polarisation basis with $p$/$s$ naming conventions often used in surface optics. The $\hat{\mathbf{e}}_p$ and $\hat{\mathbf{e}}_s$ unit vectors correspond to the $\theta$ and $\phi$ unit vectors in spherical coordinates respectively. Details of this polarisation basis decomposition are provided in the Supplementary Information in Section 4. $\mathcal{A}$ then splits into $A_p(\mathbf{k}) \, \hat{\mathbf{e}}_p + A_s(\mathbf{k}) \, \hat{\mathbf{e}}_s$ and Eq.~(\ref{eq:angularspectrum}) becomes,
\begin{equation}\label{eq:psbasis}
    \mathbf{E}(\mathbf{r}) = 
    \iint
    \Big[A_p(\mathbf{k}) \, \underbrace{\hat{\mathbf{e}}_p \, e^{i \, \mathbf{k} \cdot \mathbf{r}}}_{\text{plane wave}} + A_s(\mathbf{k}) \, \underbrace{\hat{\mathbf{e}}_s \, e^{i \, \mathbf{k} \cdot \mathbf{r}}}_{\text{plane wave}}\Big] \, e^{-i \mathbf{k} \cdot \mathbf{r}_0} \, \text{sin}(\theta) \, \text{d}\theta \, \text{d}\phi, 
\end{equation}
The scalar functions $A_p$ and $A_s$ are derived from $\mathcal{A}$ via dot products such that $A_p = \mathcal{A} \cdot \hat{\mathbf{e}}_p$ and $A_s = \mathcal{A} \cdot \hat{\mathbf{e}}_s $. Eq.~(\ref{eq:psbasis}) explicitly highlights which terms correspond to plane waves of each polarisation state because these are the terms that PWS fields can replace via substitution to incorporate particle scattering into the beam fields $\mathbf{E}(\mathbf{r})$. That is to say, $\mathbf{E}(\mathbf{r})$ transitions from the free-space beam to a beam incident on a scattering structure when $\hat{\mathbf{e}}_{p,s} \, e^{i \, \mathbf{k} \cdot (\mathbf{r}-\mathbf{r}_0)} \to \mathbf{E}_{p,s}^{\text{PWS}}(\mathbf{r})$. The vector field $\mathbf{E}_{p,s}^{\text{PWS}}(\mathbf{r})$ varies with $\theta$ and $\phi$ because it represents the material structure being illuminated by a plane wave of all possible orientations. Symmetry arguments can then be applied to $\mathbf{E}_{p,s}^{\text{PWS}}(\mathbf{r})$ to greatly reduce the number of simulations required and instead implement a series of computationally cheap 3D rotations to map a small number of simulations to all possible $\mathbf{E}_{p,s}^{\text{PWS}}(\mathbf{r})$.  

The final step before numerical implementation is to discretise the continuous integrals over $\theta$ and $\phi$ into finite sums, and we arrive at the fundamental expression for numerically generating the fields of an arbitrarily structured beam incident on a material structure,
\begin{equation}\label{eq:combinePWS}
    \mathbf{E}(\mathbf{r}) \approx \sum_{p,s} \sum_\theta \sum_\phi A_{p,s}(\mathbf{k}) \, \mathbf{E}_{p,s}^{\text{PWS}}(\mathbf{r}) \, e^{-i \, \mathbf{k} \cdot \mathbf{r}_0} \, \text{sin}(\theta) \, \Delta \theta \, \Delta \phi. 
\end{equation}
The first of the three summations indicate a sum over the two polarisation states $p$ and $s$. The latter two sum over the plane wave orientations in spherical coordinates. 

In some cases, it is desirable to simulate the scenario where the symmetric particle is not located at the beam's focus. A common example of this would be for doughnut beams where the focus has a low electric field intensity. In this case, optical forces due to the electric field intensity are likely stronger away from the focus of the beam. The $e^{-i \, \mathbf{k} \cdot \mathbf{r}_0}$ term allows for the translation of the beam's focus without the need for recalculations of $\mathbf{E}_{p,s}^{\text{PWS}}(\mathbf{r})$ by changing the relative phases of different PWS components. This allows for the efficient creation of informative graphs such as a force maps, in which the optical force is plotted as a function of the particle position within the beam, in a post-processing manner. This is demonstrated in Section \ref{subsec:cone}.

\subsection{Angular spectrum of non-paraxial beams}

One of the key features of this method lies in its ability to rapidly compute a wide range of illumination types from the same numerical simulation data files. In the case where a numerical solver is capable of illuminating an object with a sophisticated beam structure (more than a single plane wave), any change in the beam parameters such as angle of incidence, focus position, beam waist and polarisation will always necessitate a completely new and potentially costly simulation. Any parameter sweeps will therefore be computationally expensive. Our method requires a minimal number of preliminary numerical simulations with plane waves and then manipulates this minimal data to achieve any illumination structure. Crucially, this allows for parameter sweeps such as angle of illumination or translating the beam's focal point to be performed efficiently by post-processing the same initial plane wave simulations.

We shall now discuss how an incident beam field can be expressed with the angular spectrum and implemented in our approach, but we emphasise that the angular spectrum representation is not limited to beam structures; any electromagnetic field incident on an object can be represented with the angular spectrum approach using Eq.~(\ref{eq:angularspectrum}). 
Optical beams are often expressed in their paraxial approximation forms to simplify calculations \cite{Saleh2007} but in doing so, one loses both an exact solution to Maxwell's equations and some physical properties related to focusing such as a longitudinal field component \cite{Bliokh2015a,Nieminen2008,Gahagan1996,Zhao2007,Eismann2020,Li2017,Liu2019}. We do not wish our method to suffer from these limitations. To rectify this, one can use the angular spectrum approach to begin with a paraxial equation and then determine the longitudinal component to create a 3D field distribution that obeys the laws of electromagnetism in all regimes and across all of space. This is possible because the angular spectrum is based on a linear combination of plane wave solutions, and each plane wave is an exact solution of Maxwell's equations. To do this, we implement the following procedure.

\subsubsection{Integrating a paraxial beam}

We begin with the fields of a paraxial beam with a propagation axis denoted by $\hat{\mathbf{K}}$. To start our argument, lets restrict the propagation to the $z$-axis (i.e. $\hat{\mathbf{K}} \parallel \hat{\mathbf{z}}$) and only require the fields in the focal plane, which we specify as $z=0$. The paraxial field is therefore denoted by $\mathbf{E}_\perp^{\hat{\mathbf{K}} \parallel \hat{\mathbf{z}}}(x,y,0)$, where the $\perp$ symbol indicates that the fields are only polarised in the plane transverse to the propagation direction. The fields of the beam may be defined in Cartesian coordinates so it is convenient to recast Eq.~(\ref{eq:angularspectrum}) into the Cartesian basis with $\mathbf{k} = (k_x,k_y,k_z)$,
\begin{equation}\label{eq:z0plane}
    \mathbf{E}_\perp^{\hat{\mathbf{K}} \parallel \hat{\mathbf{z}}}(x,y,0) = \underset{k_x^2+k_y^2 \leq k^2}{\iint}\mathcal{F}_\perp^{\hat{\mathbf{K}} \parallel \hat{\mathbf{z}}}(k_x,k_y) \, e^{i(k_x \, x + k_y \, y)} \, \text{d}k_x \, \text{d}k_y.
\end{equation}
Here $\mathcal{F}$ denotes the angular spectrum, but defined on the 2D $z=0$ plane rather than a sphere dictated by $\theta$ and $\phi$ (refer to the Supplementary Information for more details).  We relate $\mathcal{F}$ to the $\mathcal{A}$ that appears in Eq.~(\ref{eq:angularspectrum}) later in this procedure. The inverse of this Fourier transform is,
\begin{equation}\label{eq:AfromE}
    \mathcal{F}_\perp^{\hat{\mathbf{K}} \parallel \hat{\mathbf{z}}}(k_x,k_y) = \frac{1}{4 \pi^2}
    \overset{\infty}{\underset{-\infty}{\iint}}
    \mathbf{E}_\perp^{\hat{\mathbf{K}} \parallel \hat{\mathbf{z}}}(x,y,0) \, e^{-i(k_x \, x + k_y \, y)} \, \text{d}x \, \text{d}y.
\end{equation}
For a simple paraxial Gaussian beam centred at the origin, $\mathbf{E}_{\perp}^{\hat{\mathbf{K}} \parallel \hat{\mathbf{z}}}(x,y,0)= (E_x,E_y,0) \, e^{-\frac{x^2+y^2}{w^2}}$, where $w$ is the beam waist, and $E_x$ and $E_y$ are complex scalars that determine the phase, amplitude and polarisation of the beam's transverse field \cite{Lakhtakia1992,Novotny2006,Rotenberg2012,Picardi2017,Kingsley-Smith2018}. The corresponding angular spectrum would be $\mathcal{F}_\perp^{\hat{\mathbf{K}} \parallel \hat{\mathbf{z}}}(k_x,k_y) = (E_x,E_y,0) \, w^2 \, e^{-\frac{w^2}{4}(k_x^2 + k_y^2)}$. This step can be streamlined by referring to the table in the Supplementary Information Section 3 which lists forms of $\mathcal{F}_\perp^{\hat{\mathbf{K}} \parallel \hat{\mathbf{z}}}$ for some common beam modes. 

\subsubsection{Extending from focal plane to all space}

To extend the fields on the $z=0$ plane from Eq.~(\ref{eq:z0plane}) to all space, we specify that the exact field $\mathbf{E}$ (which we wish to compute in the end) must fulfil the wave equation $\mathbf{k} \cdot \mathbf{k} = k^2$ and consists of only propagating components ($k_x^2+k_y^2 \leq k^2$). This leads to the propagator $e^{i k_z z}$ where $k_z = \sqrt{k^2-k_x^2-k_y^2}$ and $k_z>0$. The field can therefore be extended to all space via this propagator so that,
\begin{equation}\label{eq:EfromF}
    \mathbf{E}_\perp^{\hat{\mathbf{K}} \parallel \hat{\mathbf{z}}}(\mathbf{r}) = \underset{k_x^2+k_y^2 \leq k^2}{\iint} \mathcal{F}_\perp^{\hat{\mathbf{K}} \parallel \hat{\mathbf{z}}}(k_x,k_y) \, e^{i \mathbf{k} \cdot \mathbf{r}} \, \text{d}k_x \, \text{d}k_y.
\end{equation}

\subsubsection{Correcting for longitudinal fields}

$\mathcal{F}_\perp^{\hat{\mathbf{K}} \parallel \hat{\mathbf{z}}}$ only has components $(\mathcal{F}_x,\mathcal{F}_y,0)$ in the transverse plane and lacks a longitudinal field component, owing to the fact that it is derived from a paraxial field. Now consider the field:
\begin{equation}\label{eq:E3DfromF3D}
    \mathbf{E}^{\hat{\mathbf{K}} \parallel \hat{\mathbf{z}}}(\mathbf{r}) = \underset{k_x^2+k_y^2 \leq k^2}{\iint} \mathcal{F}^{\hat{\mathbf{K}} \parallel \hat{\mathbf{z}}}(k_x,k_y) \, e^{i \mathbf{k} \cdot \mathbf{r}} \, \text{d}k_x \, \text{d}k_y.
\end{equation}
This is the full true Maxwell field including longitudinal components, with no approximation, and its only condition being that its transverse components match $\mathbf{E}^{\hat{\mathbf{K}} \parallel \hat{\mathbf{z}}}_\perp(\mathbf{r})$. Explicitly, $\mathcal{F} = \mathcal{F}_x \hat{\mathbf{x}} + \mathcal{F}_y \hat{\mathbf{y}} + \mathcal{F}_z \hat{\mathbf{z}}$ whilst $\mathcal{F}_\perp = \mathcal{F}_x \hat{\mathbf{x}} + \mathcal{F}_y \hat{\mathbf{y}}$. The missing component can be retrieved by invoking Maxwell's equation $\mathbf{\nabla} \cdot \mathbf{E} = 0$ and applying it to Eq.~(\ref{eq:E3DfromF3D}) to obtain the condition $\mathbf{k} \cdot \mathbf{E} = 0$. This condition explicitly states that $k_x \, \mathcal{F}_x + k_y \, \mathcal{F}_y + k_z \, \mathcal{F}_z = 0$ and so,
\begin{equation}\label{eq:Az}
    \mathcal{F}_z = -\frac{\mathcal{F}_x \, k_x + \mathcal{F}_y \, k_y}{k_z}.
\end{equation}
This is to say that when the fields in the transverse plane are known, the longitudinal field can be easily reconstructed using this simple equation and turns $\mathcal{F}_\perp^{\hat{\mathbf{K}} \parallel \hat{\mathbf{z}}} \to \mathcal{F}^{\hat{\mathbf{K}} \parallel \hat{\mathbf{z}}} $. 
For our Gaussian beam example, this results in $\mathcal{F}^{\hat{\mathbf{K}} \parallel \hat{\mathbf{z}}}(k_x,k_y) = (E_x,E_y,E_z) \, w^2 \, e^{-\frac{w^2}{4}(k_x^2 + k_y^2)}$, where $E_z = -\frac{E_x \, k_x + E_y \, k_y}{k_z}$. 

\subsubsection{Defining a piecewise angular spectrum}

Since Eq.~(\ref{eq:psbasis}) computes the integration of $\mathcal{A}$ over the surface of a 3D sphere and Eq.~(\ref{eq:E3DfromF3D}) integrates $\mathcal{F}$ over the 2D $z=0$ plane when $k_z>0$, $\mathcal{F}$ must be related to $\mathcal{A}$ with a piecewise mapping and accounting for the different Jacobians in the integrals. The differential solid angle $\text{d}\Omega = \text{sin}(\theta) \, \text{d}\theta \, \text{d}\phi = \frac{1}{k \, k_z} \, \text{d}k_x \, \text{d}k_y$ \cite{Novotny2006} and results in,
\begin{equation}
    \mathcal{A}^{\hat{\mathbf{K}} \parallel \hat{\mathbf{z}}}(\mathbf{k}) = 
    \begin{cases} 
      k \, k_z \, \mathcal{F}^{\hat{\mathbf{K}} \parallel \hat{\mathbf{z}}}(k_x,k_y), & \text{if} \, k_z > 0 \\
      0, & \text{if} \, k_z \leq 0 
    \end{cases}
\end{equation}
For our simple Gaussian beam example, this indicates that $\mathcal{A}^{\hat{\mathbf{K}} \parallel \hat{\mathbf{z}}}(\mathbf{k}) = (E_x,E_y,E_z) \, k \, k_z \, w^2 \, e^{-\frac{w^2}{4}(k_x^2 + k_y^2)}$. 

\subsubsection{Rotation of beam axis}

Eq.~(\ref{eq:AfromE}) requires the paraxial beam's propagation to be parallel to the $z$-axis, otherwise the integral will not converge. However, $\mathcal{A}^{\hat{\mathbf{K}} \parallel \hat{\mathbf{z}}}$ can be mapped to any arbitrarily orientated but otherwise similar beam by use of conventional 3D rotation matrices (see Supplementary Information section 2 for details). This is the step where $\mathcal{A}$ becomes more convenient than $\mathcal{F}$ because it is defined on the surface of a sphere and so rotates easily without the need for additional corrections. In this way, the beam axis and polarisation state can be changed, via simple rotations, to match whatever configuration is desired, such that $\mathcal{A}^{\hat{\mathbf{K}} \parallel \hat{\mathbf{z}}} \to \mathcal{A}$. 

\subsubsection{Obtaining non-paraxial 3D fields}

The rotated angular spectrum $\mathcal{A}$ can now be decomposed into $A_{p,s}$ in the manner described in Sec.~\ref{subsec:addstructure} and Eq.~(\ref{eq:combinePWS}) is used to generate the fields of the desired beam, complete with any inserted particle scattering. 
See the Supplementary Information for more details. At this stage, one can now obtain the fields of a non-paraxial beam $\mathbf{E}(\mathbf{r})$ with any $\hat{\mathbf{K}}$ by simply starting with  $\mathbf{E}_\perp^{\hat{\mathbf{K}} \parallel \hat{\mathbf{z}}}(x,y,0)$.
For our Gaussian beam example propagating along $\hat{\mathbf{z}}$, the non-paraxial electric field takes the form $\mathbf{E}(\mathbf{r}) = \int^{2\pi}_{\phi=0}\int^{\pi/2}_{\theta=0} 
(E_x,E_y,E_z) \, k \, k_z \, w^2 \, e^{-\frac{w^2}{4}(k_x^2 + k_y^2)}
\, e^{i \, \mathbf{k} \cdot (\mathbf{r}-\mathbf{r}_0)} \, \text{sin}(\theta) \, \text{d}\theta \, \text{d}\phi$, where $(k_x,k_y,k_z) \equiv  k\big(\text{sin}(\theta) \, \text{cos}(\phi),\text{sin}(\theta) \, \text{sin}(\phi), \text{cos}(\theta)\big)$. 

We note that whilst we focus on the electric field in this paper, the same method applies to the magnetic field, and magnetic components of structured beams are calculated and used in the Results section.

\subsection{Optical force and torque}

When a beam is incident on a particle, it can manipulate its position and orientation via optical forces and optical torques. These phenomena are driven by the transfer of linear and angular momentum from the beam to the particle respectively, and can be readily calculated from 3D electromagnetic field data via the Maxwell stress tensor (MST) method. The MST, denoted by $\overset\leftrightarrow{\mathbf{T}}$, represents the flow of momentum in the field and can be expressed as \cite{Novotny2006},
\begin{equation}\label{eq:mst}
    \langle\overset\leftrightarrow{\mathbf{T}}\rangle = \frac{1}{2} \mathfrak{Re} \bigg\{\varepsilon \mathbf{E} \otimes \mathbf{E}^* + \mu \mathbf{H} \otimes \mathbf{H}^* - \frac{1}{2} \big(\varepsilon |\mathbf{E}|^2 + \mu |\mathbf{H}|^2\big)\overset\leftrightarrow{\mathbf{I}}\bigg\},
\end{equation}
where $\mathbf{E}$ and $\mathbf{H}$ are the total electric and magnetic fields, $\otimes$ denotes the outer product of two vectors, asterisks represent complex conjugations, $\overset\leftrightarrow{\mathbf{I}}$ is the $3 \times 3$ identity matrix and $\varepsilon$ and $\mu$ are the permittivity and permeability of the medium, respectively. The angular brackets indicate a time-averaged quantity. A force occurs when there is a net inward or outward flow of momentum into the body experiencing the force and can be quantified by the flux surface integral \cite{Novotny2006},
\begin{equation}\label{eq:force}
    \langle\mathbf{F}\rangle = \int_{S}^{} \langle\overset\leftrightarrow{\mathbf{T}}\rangle \cdot \textbf{\^n} \, \text{d}S,
\end{equation}
where $\mathbf{F}$ is the optical force and $\textbf{\^n}$ is the outward normal vector of any arbitrary closed surface $S$ enclosing the body. Likewise, the torque can be calculated by imposing the definition of angular momentum with a cross product such that,
\begin{equation}\label{eq:torque}
    \langle\mathbf{N}\rangle = \int_{S} \Big(\mathbf{r} \times \big\langle\overset\leftrightarrow{\mathbf{T}}\big\rangle \Big) \cdot \hat{\textbf{n}} \, \text{d}S,
\end{equation}
where $\mathbf{N}$ is the optical torque and $\mathbf{r}$ is the position vector from the axis of rotation. The collective cross product term in the brackets is sometimes referred to as the angular momentum flux tensor \cite{Barnett2002,Chen2015}. With these expressions, one can immediately determine particle dynamics originating from electromagnetism regardless of the particle's properties. The only quantities required to calculate forces and torques from Eqs.~(\ref{eq:force}) and (\ref{eq:torque}) are the full electromagnetic fields (both electric and magnetic) of the incident beam scattering on the particle, which our method provides. 

\section{Results}

The strength of our approach lies in its high degree of generality and applicability. To demonstrate this, we provide analysis of three distinct nanophotonic systems. The PWS were conducted in \textit{CST Microwave Studio}. 

\subsection{Core-shell particle in focused Gaussian beam}

\begin{figure}[t]
    \centering
    \includegraphics[scale=1]{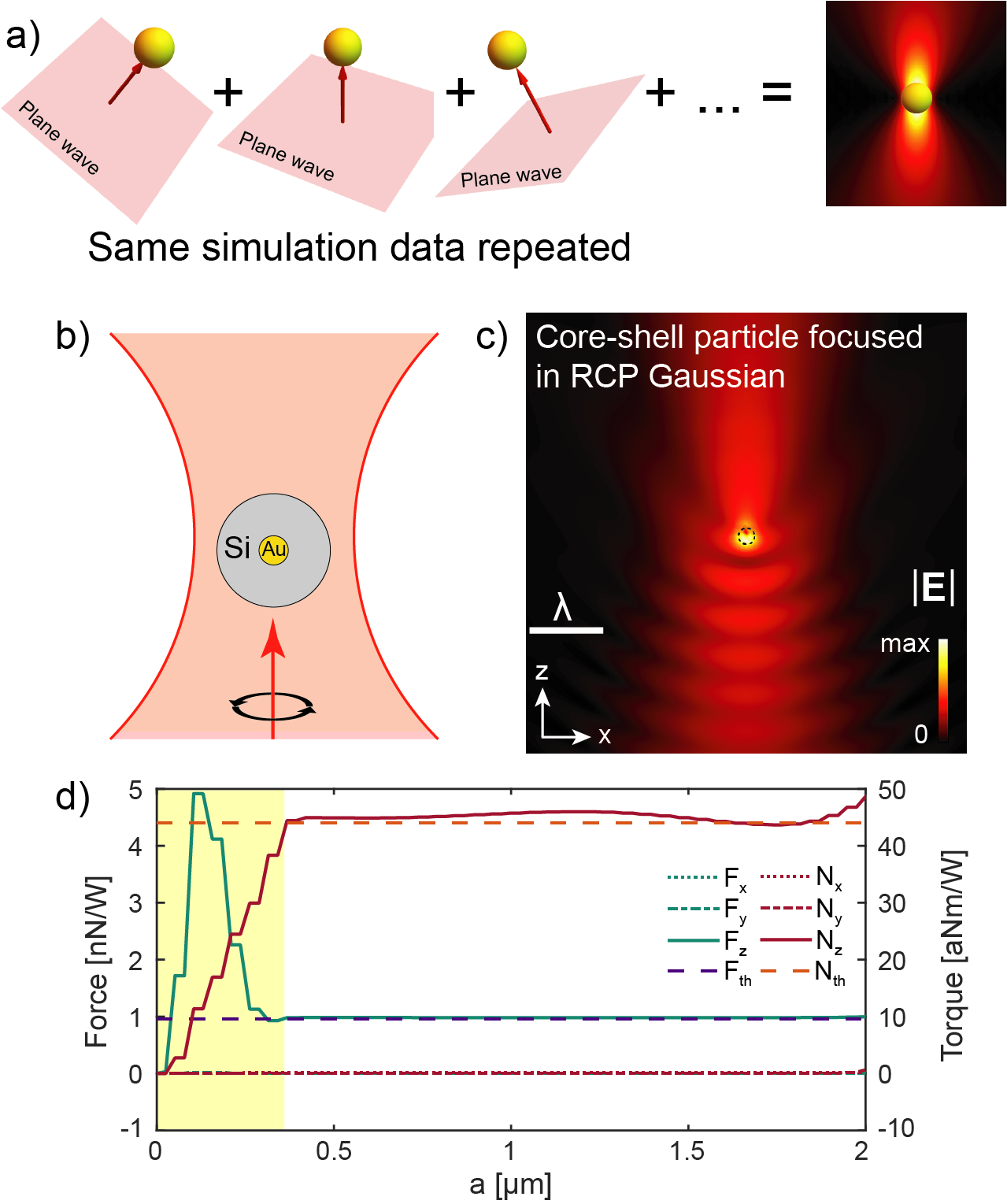}
    \caption[The field distribution, force and torque on a core-shell particle within a focused Gaussian beam]{a) By incorporating plane wave simulation data into the angular spectrum and utilising the symmetry of the system, the scattering of a particle in an arbitrarily structured beam can be calculated from a single simulation data set. b) A gold-core silicon-shell particle illuminated with a focused Gaussian beam with right-hand circular polarisation (RCP). Diagram not to scale.  c) A cross-section of $|\mathbf{E}|$ through the beam and particle in the $y=0$ plane. The outer boundary of the particle is indicated with a black dotted line. d) The time-averaged force and torque on the particle located at the focus as a function of the integration surface size $a$, normalised by the beam power. A flat line indicates a stable and reliable result. The yellow region indicates when the integration surface intersects with the particle, and so these results are invalid and ignored. The purple and orange dashed lines indicate the theoretical values for the force and torque, respectively, calculated via Mie theory.}
    \label{fig:coreshell}
\end{figure}

Multilayered spherical particles are a common solution when a particle's scattering/absorption needs to be engineered. Such particles can be fabricated with modern nanofabrication techniques \cite{Wang2018} and have been proposed for applications including photocatalysis, nanolasers, spontaneous emission enhancement and as a basis for nonlinear phenomena. In this example, we use the gold-core silicon-shell design outlined by Feng et. al. \cite{Feng2017} which exhibits a pure magnetic dipole resonance at a wavelength of $\lambda = 1.3 \, \mu m$. This core-shell particle therefore probes the magnetic field of any incident beam. The core radius is 62 nm and the shell radius is 180 nm. The rotational symmetry of this geometry means that only two PWS are needed (one for each polarisation) in order to calculate the interaction between a multilayered sphere and a beam of any structure. 

Fig. \ref{fig:coreshell}b shows a schematic of the system with the beam propagating along the positive $z$-axis. The beam waist $w$ is set to $0.5 \lambda$ where $\lambda = 1.3 \, \mu m$, and the surrounding medium has $n=1$. After a very quick simulation of two plane wave simulations scattering on the particle in CST Microwave Studio, our post-processing code was applied to them. The output is the full fields for an incident circularly polarised tightly focused Gaussian beam scattering on the core-shell particle. The angular spectrum of this illumination is conveniently provided by the package's analytical workbook. Fig. \ref{fig:coreshell}c shows a cross section of $|\mathbf{E}|$ in the $y=0$ plane and the particle is outlined with a black dotted line. The interference between the incident beam and the particle's scattering is clearly visible in the bottom half of the image. This 3D electromagnetic field distribution was used to calculate the optical force and torque on the particle using the MST method.
A cubic surface enclosing the particle was chosen for the integration. In principle, any size of this cube enclosing the particle should result in the same force and torque. Numerically, however, there are always variations, so it is important to check consistency and obtain a statistically accurate value for these quantities by varying the size of the integration surface (see Fig. \ref{fig:coreshell}d). Fig. \ref{fig:coreshell}d shows the Cartesian components of the force and torque calculated across a range of integration cube sizes, and flat lines indicate a reliable result. The yellow region ($a<360$ nm) indicates where the integration surface is too small and intersects with the particle, resulting in inaccurate results. 

A spherical particle is a special case in scattering problems because Mie theory provides an exact analytical solution to the problem. Therefore, we can evaluate the accuracy of our numerical method by comparing with the results obtained through Mie theory. The details of these analytical calculations are provided in the Supplementary Information, and the theoretical values for the force and torque (denoted by $F_\text{th}$ and $N_\text{th}$) are plotted in Fig. \ref{fig:coreshell}d with blue and orange dotted lines, respectively. We observe a strong agreement between theoretical and numerical results, indicating a reliable numerical calculation. 

As described in detail above, we could now vary any parameter of this beam (waist, size, focus location, type of beam, etc) and simply run the post-processing steps on the same two PWS to obtain the new fields, force and torque. 

\subsection{Plasmonic nanocone in azimuthally polarised beam}\label{subsec:cone}

The next phase in this demonstration is the use of sophisticated structures that are more difficult to solve. Without spherical symmetry, analytical methods such as Mie theory no longer provide an immediate answer to the scattering problem. Cylinders are commonplace in nanophotonics \cite{Wang2019a} but cylindrical symmetry applies to a far wider range of geometries than just cylinders; structures such as cones \cite{Cordova-Castro2019,Evlyukhin2011}, tori/rings \cite{Aizpurua2003,Mary2007,Dutta2008,Feng2014}, tubes and core-shell cylinders \cite{DeAngelis2013} are examples of this. 
Out of these structures, a conical geometry is perhaps one of the most difficult to calculate because it lacks the cross-sectional mirror symmetry that would enable a further reduction in the number of PWS from $0 \leq \theta \leq \pi$ to $0 \leq \theta \leq \frac{\pi}{2}$. For this reason, the next example uses a nanocone. 

Gold nanocones are already experimentally viable and the plasmonic nature enables stronger scattering \cite{Cordova-Castro2019}. The illuminating beam's complexity has also been increased by selecting a focused azimuthally polarised vortex beam to emphasise the range of possible beam options. Furthermore, the angle of incidence is set to 45$^\circ$ to highlight that the beam axis can be chosen independently of the cylindrical symmetry axis. The wavelength is set to 532 nm to match the plasmonic resonance range of gold, and the beam waist is set to 0.8 $\lambda$.
Once again, PWS were carried out in CST Microwave Studio for plane waves incident on the nanocone at different angles of incidence $\theta$ (but always keeping $\phi = 0$). This PWS data was then post-processed using our code to obtain the full field distributions under the desired incidence. 

\begin{figure}
    \centering
    \includegraphics[scale=1]{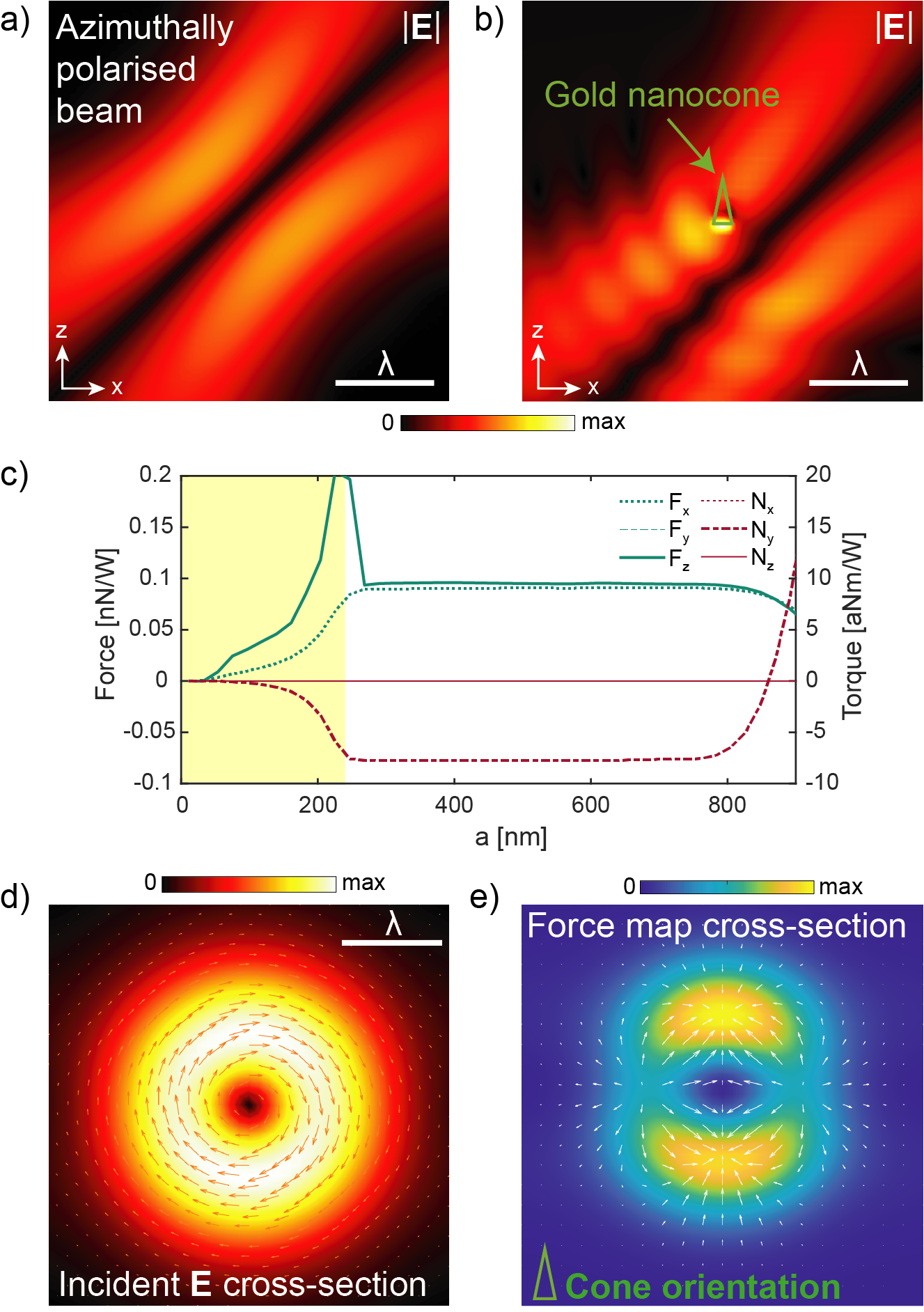}
    \caption[The field distribution, force and torque on a gold nanocone within a tilted azimuthally polarised beam]{a) An azimuthally polarised beam propagating in the $(1,0,1)$ direction in free space with $\lambda = 532$ nm and $w = 0.8 \lambda$. b) The same azimuthally polarised beam is translated 300 nm down and 150 nm to the right and the gold nanocone is introduced. Its height is 240 nm and the base radius is 50 nm. c) The optical force and torque in b) is plotted against the MST integration cube's side length to confirm a stable result. The yellow region indicates where the integration cube intersects with the nanocone, and these results can be neglected. d) A cross section of the azimuthally polarised beam in a) depicting the electric field polarisation. e) A force map of the gold nanocone in the same cross section plane as d). The arrows indicate the components of the time-averaged force in the cross section plane, and the colours represent the total magnitude of the force. }
    \label{fig:nanocone}
\end{figure}

Fig. \ref{fig:nanocone}a shows the $y=0$ cross-section plane of the beam's electric field distribution in free-space and Fig. \ref{fig:nanocone}b introduces the gold nanocone to the beam at the point of highest intensity. The nanocone has a height of 240 nm and a maximum radius of 50 nm.  Similar to Fig. \ref{fig:coreshell}b, we clearly observe the nanocone's scattering interfering with the incident beam and further see a hot spot at the base of the cone. This hot spot agrees with previous results in literature \cite{Cordova-Castro2019} where a similar excitation is seen with the same material and a similar wavelength. Fig. \ref{fig:nanocone}c shows the optical force and torque on the nanocone and is analogous to Fig. \ref{fig:coreshell}c. The force is directed along the beam's propagating direction so can be explained as a simple radiation pressure on the nanocone. $F_z$ is also slightly larger than $F_x$, suggesting that the nanocone's hot spot is being pushed towards the maximum of the beam. The torque along the $-y$ direction indicates that the nanocone will rotate anti-clockwise from Fig. \ref{fig:nanocone}b's perspective. Further iterative calculations could be conducted in order to determine an orientation with rotational equilibrium. 

Finally, in order to highlight the method's power and without requiring additional plane wave simulations, we applied the post-processing code to sweep the nanocone position across the beam by changing $\mathbf{r}_0$ in Eq.~(\ref{eq:combinePWS}) and calculated a force map shown in Fig. \ref{fig:nanocone}e. Each arrow in this figure represents a full post-processing simulation each including MST integration in a cube around the particle. Producing a force map would have been a tedious process if performing individual numerical simulations for each nanocone position, but we can use our approach to create such a plot in a matter of minutes on a standard desktop computer, all from the same set of PWS.

\subsection{Vortex beam reflection}

The previous examples focused on standalone particles, but this calculation method is equally valid for configurations involving planar structures. A planar interface is naturally axisymmetric and so rotations around the surface's normal axis yield physically identical results and therefore can be exploited to reduce the dimensionality of the problem. A multilayered or stratified structure is merely an extension of this problem and can be calculated in a similar manner to the previous examples by also applying the well-established transfer matrix method \cite{Born1980}. 


\begin{figure}[b]
    \centering
    \includegraphics[scale=1]{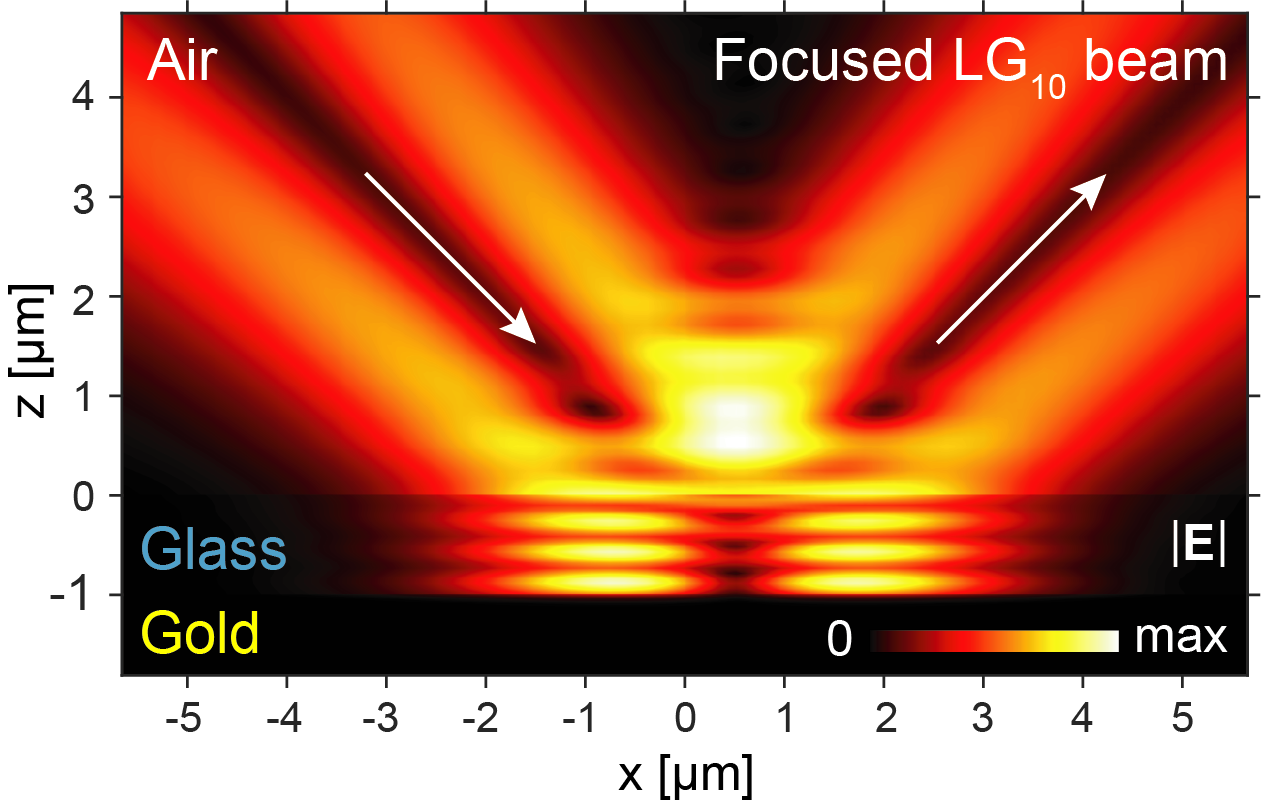}
    \caption{The electric field distribution of a \textit{p}-polarised Laguerre-Gaussian vortex beam incident at 45$^\circ$ focused onto a glass slab with a thick gold substrate, with $\lambda=808$ nm and $w=0.75\lambda$. The white arrows indicate the incident and reflected beam propagation directions.}
    \label{fig:reflectedvortex}
\end{figure}

We demonstrate this case with a \textit{p}-polarised focused vortex beam (Laguerre-Gaussian with $l=1$ and $m=0$) reflecting off of a glass-gold multilayered structure, and the resulting electric field distribution is portrayed in Fig. \ref{fig:reflectedvortex}. The beam has a wavelength of 808 nm and is focused at the origin to a beam waist of 0.75$\lambda$. The glass, with a refractive index of 1.5, is 1 $\mu$m thick and semi-infinite gold \cite{Johnson1972} is placed beneath it. The incident polarisation singularity is clearly visible along the $z=-x$ line and an interference pattern is observed around the air-glass interface. 

The 3D electromagnetic fields were calculated in Fig. \ref{fig:reflectedvortex} using both a commercial frequency-domain numerical solver with periodic Floquet boundary conditions and with an analytical calculation of the multilayer scattering of plane waves. The latter is naturally faster computationally but is limited to homogeneous layered systems. Numerical solvers enable the calculation of more complicated axisymmetric structures such as an isolated nanorod or nanocone on a substrate, or a recess like an isolated nanopore. Whilst these structures are regularly found within tightly packed periodic metamaterials, the isolated structure results still provide valuable information about the structure's resonant behaviour under complex beam illumination, and how the substrate augments these properties. In this way, we expect our package to prove a valuable tool for experimental groups looking to gain a deeper understanding of their experimental nanomaterials that consist of these elements when illuminated with a complex beam. In the event where the rod/cone/pore separations are large enough for neighbouring element interactions to become negligible, this axisymmetric angular spectrum approach will yield approximate quantitative results.


\section{Discussion}

This paper proposes a theoretical beam-generation method which strength lies in its generality, and its applicability extends beyond the cases discussed here. One such example is that of optical pulses. Our examples are all monochromatic but this is for the sake of simplicity, rather than evidence of a limitation. To extend to pulses, the angular spectrum of the pulse must include a further integral over the temporal frequency. The frequency spectrum of the pulse can then be inserted into the integrand and the equations for the electromagnetic fields are equally valid. 

The demonstrations in the previous section only incorporate a single propagating beam for the illumination of a target object, but nothing prevents additional beams being added to these systems. Multiple incident beams can be calculated either by adding the fields of single beam systems together or by combining the angular spectra of each beam. Areas such as levitating optomechanics regularly implement counter-propagating beams to trap particles inside of a standing wave \cite{Millen2020,Kuhn2017,Stickler2018}. These types of systems can be analysed in detail with our method. Likewise, optical lattice systems can be calculated by defining a discrete symmetry in the angular spectra. 

Lastly, this method is innately suited to anisotropic systems. Figs. \ref{fig:nanocone} and \ref{fig:reflectedvortex} illustrate cases where the geometry of scattering object is anisotropic, whereby the material dimensions extend into each Cartesian direction differently. However, one can also consider cases where the intrinsic material properties, such as the refractive index, are fundamentally anisotropic as long as the material anisotropy respects the rotational symmetry around the $z$-axis. In this case, even a spherical particle can become a difficult problem. The combination of geometrical anisotropy and material anisotropy would invoke sophisticated near-field and far-field scattering and may be of interest in some areas of optics. 

The interaction between electromagnetic fields and matter is at the heart of countless technologies and the full-wave 3D calculation of these fields is a constant problem in photonics. We have demonstrated a complete method for computing the interaction between an arbitrarily complicated incident beam and a rotationally symmetric structure with a high degree of generality, requiring only the prior simulation of plane wave incidence, which is a simple and fast calculation that can be performed by all electromagnetic solver packages. There is no need for additional approximations in the tightly focused limit and vastly increases the range of systems that can be calculated with numerical solvers. Subwavelength and near-field scattering is accurately constructed and electromagnetic quantities such as optical forces and torques can be reliably calculated with the resultant complex field data. Many structures used in nanophotonics today, including spheres, rods, tubes and cones, fall within the capabilities of this method and thus a many types of analysis can be conducted on said structures with a minimal number and efficient use of numerical solvers. We believe that this will prove a valuable tool for many researchers and so provide an open-source software package called \textit{BEAMS} (Beams scattering through Electromagnetic Axisymmetric Multilayers and Structures) where this post-processing method is implemented in MATLAB and provided at Ref. \cite{BEAMS2023}.

\section*{Acknowledgements}

J. J. Kingsley-Smith thanks Dr M. F. Picardi for helpful discussions over software design. This work was supported by the European Research Council Starting Grant ERC-2016-STG-714151-PSINFONI. 

\section*{References}
\bibliography{references}

\begin{thebibliography}{10}

\bibitem{Abbott2016}
B.~P. Abbott~\textit{et. al.}
\newblock {Observation of Gravitational Waves from a Binary Black Hole Merger}.
\newblock {\em Physical Review Letters}, 116(6):061102, feb 2016.

\bibitem{Moffitt2008}
Jeffrey~R. Moffitt, Yann~R. Chemla, Steven~B. Smith, and Carlos Bustamante.
\newblock {Recent Advances in Optical Tweezers}.
\newblock {\em Annual Review of Biochemistry}, 77(1):205--228, jun 2008.

\bibitem{Righini2009}
M.~Righini, P.~Ghenuche, S.~Cherukulappurath, V.~Myroshnychenko, F.~J.
  {Garc\'{i}a de Abajo}, and R.~Quidant.
\newblock {Nano-optical Trapping of Rayleigh Particles and Escherichia coli
  Bacteria with Resonant Optical Antennas}.
\newblock {\em Nano Letters}, 9(10):3387--3391, oct 2009.

\bibitem{Brown2011}
Alice C.~N. Brown, Stephane Oddos, Ian~M. Dobbie, Juha-Matti Alakoskela,
  Richard~M. Parton, Philipp Eissmann, Mark A.~A. Neil, Christopher Dunsby,
  Paul M.~W. French, Ilan Davis, and Daniel~M. Davis.
\newblock {Remodelling of Cortical Actin Where Lytic Granules Dock at Natural
  Killer Cell Immune Synapses Revealed by Super-Resolution Microscopy}.
\newblock {\em PLoS Biology}, 9(9):e1001152, sep 2011.

\bibitem{Svoboda1994}
K~Svoboda and S~M Block.
\newblock {Biological Applications of Optical Forces}.
\newblock {\em Annual Review of Biophysics and Biomolecular Structure},
  23(1):247--285, jun 1994.

\bibitem{Kuo2001}
Scot~C. Kuo.
\newblock {Using Optics to Measure Biological Forces and Mechanics}.
\newblock {\em Traffic}, 2(11):757--763, nov 2001.

\bibitem{Juan2011}
Mathieu~L Juan, Maurizio Righini, and Romain Quidant.
\newblock {Plasmon nano-optical tweezers}.
\newblock {\em Nature Photonics}, 5(6):349--356, jun 2011.

\bibitem{Greulich2017}
Karl~Otto Greulich.
\newblock {Manipulation of cells with laser microbeam scissors and optical
  tweezers: a review}.
\newblock {\em Reports on Progress in Physics}, 80(2):026601, feb 2017.

\bibitem{Xin2019}
Hongbao Xin and Baojun Li.
\newblock {Fiber-based optical trapping and manipulation}.
\newblock {\em Frontiers of Optoelectronics}, 12(1):97--110, mar 2019.

\bibitem{Bustamante2020}
Carlos Bustamante, Lisa Alexander, Kevin Maciuba, and Christian~M. Kaiser.
\newblock {Single-Molecule Studies of Protein Folding with Optical Tweezers}.
\newblock {\em Annual Review of Biochemistry}, 89(1):443--470, jun 2020.

\bibitem{Grier2003}
David~G. Grier.
\newblock {A revolution in optical manipulation}.
\newblock {\em Nature}, 424(6950):810--816, aug 2003.

\bibitem{Ashkin2000}
A.~Ashkin.
\newblock {History of optical trapping and manipulation of small-neutral
  particle, atoms, and molecules}.
\newblock {\em IEEE Journal of Selected Topics in Quantum Electronics},
  6(6):841--856, nov 2000.

\bibitem{Marago2013}
Onofrio~M. Marag{\`{o}}, Philip~H. Jones, Pietro~G. Gucciardi, Giovanni Volpe,
  and Andrea~C. Ferrari.
\newblock {Optical trapping and manipulation of nanostructures}.
\newblock {\em Nature Nanotechnology}, 8(11):807--819, nov 2013.

\bibitem{Ren2021}
Yatao Ren, Qin Chen, Mingjian He, Xiangzhi Zhang, Hong Qi, and Yuying Yan.
\newblock {Plasmonic Optical Tweezers for Particle Manipulation: Principles,
  Methods, and Applications}.
\newblock {\em ACS Nano}, 15(4):6105--6128, apr 2021.

\bibitem{Guo2013}
HongLian Guo and ZhiYuan Li.
\newblock {Optical tweezers technique and its applications}.
\newblock {\em Science China Physics, Mechanics and Astronomy},
  56(12):2351--2360, dec 2013.

\bibitem{Babiker2019}
Mohamed Babiker, David~L Andrews, and Vassilis~E Lembessis.
\newblock {Atoms in complex twisted light}.
\newblock {\em Journal of Optics}, 21(1):013001, jan 2019.

\bibitem{Allen1999}
L.~Allen, M.J. Padgett, and M.~Babiker.
\newblock {IV The Orbital Angular Momentum of Light}.
\newblock In {\em Progress in Optics}, chapter~4, pages 291--372. Elsevier,
  1999.

\bibitem{Yao2011}
Alison~M. Yao and Miles~J. Padgett.
\newblock {Orbital angular momentum: origins, behavior and applications}.
\newblock {\em Advances in Optics and Photonics}, 3(2):161, jun 2011.

\bibitem{Curtis2003}
Jennifer~E. Curtis and David~G. Grier.
\newblock {Structure of Optical Vortices}.
\newblock {\em Physical Review Letters}, 90(13):133901, apr 2003.

\bibitem{Padgett2011}
Miles Padgett and Richard Bowman.
\newblock {Tweezers with a twist}.
\newblock {\em Nature Photonics}, 5(6):343--348, jun 2011.

\bibitem{Shen2019}
Yijie Shen, Xuejiao Wang, Zhenwei Xie, Changjun Min, Xing Fu, Qiang Liu, Mali
  Gong, and Xiaocong Yuan.
\newblock {Optical vortices 30 years on: OAM manipulation from topological
  charge to multiple singularities}.
\newblock {\em Light: Science \& Applications}, 8(1):90, dec 2019.

\bibitem{Zhan2009}
Qiwen Zhan.
\newblock {Cylindrical vector beams: from mathematical concepts to
  applications}.
\newblock {\em Advances in Optics and Photonics}, 1(1):1, jan 2009.

\bibitem{Hanifeh2020}
Mina Hanifeh, Mohammad Albooyeh, and Filippo Capolino.
\newblock {Optimally Chiral Light: Upper Bound of Helicity Density of
  Structured Light for Chirality Detection of Matter at Nanoscale}.
\newblock {\em ACS Photonics}, page acsphotonics.0c00304, sep 2020.

\bibitem{Wang2018b}
Xuewen Wang, Zhongquan Nie, Yao Liang, Jian Wang, Tao Li, and Baohua Jia.
\newblock {Recent advances on optical vortex generation}.
\newblock {\em Nanophotonics}, 7(9):1533--1556, aug 2018.

\bibitem{Wang2019}
Peipei Wang, Junmin Liu, Yanliang He, Jun Liu, Xinxing Zhou, Huapeng Ye, Ying
  Li, Shuqing Chen, and Dianyuan Fan.
\newblock {Arbitrary Cylindrical Vector Beam Generation Using Cross-Polarized
  Modulation}.
\newblock {\em IEEE Photonics Technology Letters}, 31(11):873--876, jun 2019.

\bibitem{Zhu2019}
Long Zhu and Jian Wang.
\newblock {A review of multiple optical vortices generation: methods and
  applications}.
\newblock {\em Frontiers of Optoelectronics}, 12(1):52--68, mar 2019.

\bibitem{Tidwell1993}
Steve~C. Tidwell, Gerald~H. Kim, and Wayne~D. Kimura.
\newblock {Efficient radially polarized laser beam generation with a double
  interferometer}.
\newblock {\em Applied Optics}, 32(27):5222, sep 1993.

\bibitem{Fu2016}
Shiyao Fu, Chunqing Gao, Tonglu Wang, Shikun Zhang, and Yanwang Zhai.
\newblock {Simultaneous generation of multiple perfect polarization vortices
  with selective spatial states in various diffraction orders}.
\newblock {\em Optics Letters}, 41(23):5454, dec 2016.

\bibitem{Cardano2012}
Filippo Cardano, Ebrahim Karimi, Sergei Slussarenko, Lorenzo Marrucci, Corrado
  de~Lisio, and Enrico Santamato.
\newblock {Polarization pattern of vector vortex beams generated by q-plates
  with different topological charges}.
\newblock {\em Applied Optics}, 51(10):C1, apr 2012.

\bibitem{He2022}
Jinna He, Mingli Wan, Xiaopeng Zhang, Shuqing Yuan, Liufang Zhang, and Junqiao
  Wang.
\newblock {Generating ultraviolet perfect vortex beams using a high-efficiency
  broadband dielectric metasurface}.
\newblock {\em Optics Express}, 30(4):4806, feb 2022.

\bibitem{Mandal2020}
Arabinda Mandal, Satyajit Maji, and Maruthi~M. Brundavanam.
\newblock {Common-path generation of stable cylindrical perfect vector vortex
  beams of arbitrary order}.
\newblock {\em Optics Communications}, 469:125807, aug 2020.

\bibitem{Lerman2010}
Gilad~M. Lerman, Liron Stern, and Uriel Levy.
\newblock {Generation and tight focusing of hybridly polarized vector beams}.
\newblock {\em Optics Express}, 18(26):27650, dec 2010.

\bibitem{Kasperczyk2015}
Mark Kasperczyk, Steven Person, Duarte Ananias, Luis~D. Carlos, and Lukas
  Novotny.
\newblock {Excitation of Magnetic Dipole Transitions at Optical Frequencies}.
\newblock {\em Physical Review Letters}, 114(16):163903, apr 2015.

\bibitem{Novotny2001}
L.~Novotny, M.~R. Beversluis, K.~S. Youngworth, and T.~G. Brown.
\newblock {Longitudinal Field Modes Probed by Single Molecules}.
\newblock {\em Physical Review Letters}, 86(23):5251--5254, jun 2001.

\bibitem{Wang2010}
Xi-Lin Wang, Jing Chen, Yongnan Li, Jianping Ding, Cheng-Shan Guo, and Hui-Tian
  Wang.
\newblock {Optical orbital angular momentum from the curl of polarization}.
\newblock {\em Physical Review Letters}, 105(25):253602, dec 2010.

\bibitem{Singh2009}
Rajesh Singh and James~W. Lillard.
\newblock {Nanoparticle-based targeted drug delivery}.
\newblock {\em Experimental and Molecular Pathology}, 86(3):215--223, jun 2009.

\bibitem{McPhillips2010}
John McPhillips, Antony Murphy, Magnus~P. Jonsson, William~R. Hendren, Ronald
  Atkinson, Fredrik H{\"{o}}{\"{o}}k, Anatoly~V. Zayats, and Robert~J. Pollard.
\newblock {High-Performance Biosensing Using Arrays of Plasmonic Nanotubes}.
\newblock {\em ACS Nano}, 4(4):2210--2216, apr 2010.

\bibitem{DeAngelis2013}
Francesco {De Angelis}, Mario Malerba, Maddalena Patrini, Ermanno Miele, Gobind
  Das, Andrea Toma, Remo~Proietti Zaccaria, and Enzo {Di Fabrizio}.
\newblock {3D Hollow Nanostructures as Building Blocks for Multifunctional
  Plasmonics}.
\newblock {\em Nano Letters}, 13(8):3553--3558, aug 2013.

\bibitem{Wong2013}
Bin~Sheng Wong, Sia~Lee Yoong, Anna Jagusiak, Tomasz Panczyk, Han~Kiat Ho,
  Wee~Han Ang, and Giorgia Pastorin.
\newblock {Carbon nanotubes for delivery of small molecule drugs}.
\newblock {\em Advanced Drug Delivery Reviews}, 65(15):1964--2015, dec 2013.

\bibitem{Sershen2000}
S.~R. Sershen, S.~L. Westcott, N.~J. Halas, and J.~L. West.
\newblock {Temperature-sensitive polymer-nanoshell composites for
  photothermally modulated drug delivery}.
\newblock {\em Journal of Biomedical Materials Research}, 51(3):293--298, jun
  2000.

\bibitem{Wang2019a}
Pan Wang, Mazhar~E. Nasir, Alexey~V. Krasavin, Wayne Dickson, Yunlu Jiang, and
  Anatoly~V. Zayats.
\newblock {Plasmonic Metamaterials for Nanochemistry and Sensing}.
\newblock {\em Accounts of Chemical Research}, 52(11):3018--3028, nov 2019.

\bibitem{Kale2014}
Matthew~J. Kale, Talin Avanesian, and Phillip Christopher.
\newblock {Direct Photocatalysis by Plasmonic Nanostructures}.
\newblock {\em ACS Catalysis}, 4(1):116--128, jan 2014.

\bibitem{Bohren1998}
Craig~F. Bohren and Donald~R. Huffman.
\newblock {\em {Absorption and Scattering of Light by Small Particles}}.
\newblock Wiley, New York, wiley prof edition, 1998.

\bibitem{VanderHulst2003}
H.~C. van~der Hulst.
\newblock {\em {Light Scattering by Small Particles}}.
\newblock Dover Publications Inc, New York, 2003.

\bibitem{Sipe1987}
J.~E. Sipe.
\newblock {New Green-function formalism for surface optics}.
\newblock {\em Journal of the Optical Society of America B}, 4(4):481, apr
  1987.

\bibitem{Novotny2006}
Lukas Novotny and Bert Hecht.
\newblock {\em {Principles of Nano-Optics}}.
\newblock Cambridge University Press, New York, 1st edition, 2006.

\bibitem{Vazquez-Lozano2019}
J.~Enrique V{\'{a}}zquez-Lozano, Alejandro Mart{\'{i}}nez, and Francisco~J.
  Rodr{\'{i}}guez-Fortu{\~{n}}o.
\newblock {Near-Field Directionality Beyond the Dipole Approximation: Electric
  Quadrupole and Higher-Order Multipole Angular Spectra}.
\newblock {\em Physical Review Applied}, 12(2):024065, aug 2019.

\bibitem{Peterson1997}
Andrew~F. Peterson, Scott~L. Ray, and Raj Mittra.
\newblock {\em {Computational Methods for Electromagnetics}}.
\newblock IEEE, 1997.

\bibitem{Saleh2007}
B.~E.~A. Saleh and M.~C. Teich.
\newblock {\em {Fundamentals of Photonics}}.
\newblock John Wiley \& Sons, Inc, Hoboken, New Jersey, 2nd edition, 2007.

\bibitem{Bliokh2015a}
K.~Y. Bliokh, F.~J. Rodr{\'{i}}guez-Fortu{\~{n}}o, F.~Nori, and A.~V. Zayats.
\newblock {Spin–orbit interactions of light}.
\newblock {\em Nature Photonics}, 9(12):796--808, dec 2015.

\bibitem{Nieminen2008}
Timo~A Nieminen, Alexander~B Stilgoe, Norman~R Heckenberg, and Halina
  Rubinsztein-Dunlop.
\newblock {Angular momentum of a strongly focused Gaussian beam}.
\newblock {\em Journal of Optics A: Pure and Applied Optics}, 10(11):115005,
  nov 2008.

\bibitem{Gahagan1996}
K.~T. Gahagan and G.~A. Swartzlander.
\newblock {Optical vortex trapping of particles}.
\newblock {\em Optics Letters}, 21(11):827, jun 1996.

\bibitem{Zhao2007}
Yiqiong Zhao, J.~Scott Edgar, Gavin D.~M. Jeffries, David McGloin, and
  Daniel~T. Chiu.
\newblock {Spin-to-Orbital Angular Momentum Conversion in a Strongly Focused
  Optical Beam}.
\newblock {\em Physical Review Letters}, 99(7):073901, aug 2007.

\bibitem{Eismann2020}
J.~S. Eismann, L.~H. Nicholls, D.~J. Roth, M.~A. Alonso, P.~Banzer, F.~J.
  Rodr{\'{i}}guez-Fortu{\~{n}}o, A.~V. Zayats, F.~Nori, and K.~Y. Bliokh.
\newblock {Transverse spinning of unpolarized light}.
\newblock {\em Nature Photonics}, 15(2):156--161, feb 2021.

\bibitem{Li2017}
Kun Li, Nathaniel~J. Hogan, Matthew~J. Kale, Naomi~J. Halas, Peter Nordlander,
  and Phillip Christopher.
\newblock {Balancing Near-Field Enhancement, Absorption, and Scattering for
  Effective Antenna–Reactor Plasmonic Photocatalysis}.
\newblock {\em Nano Letters}, 17(6):3710--3717, jun 2017.

\bibitem{Liu2019}
Xingguang Liu, Junqing Li, Qiang Zhang, and Mamo~Gebeyehu Dirbeba.
\newblock {Separation of chiral enantiomers by optical force and torque induced
  by tightly focused vector polarized hollow beams}.
\newblock {\em Physical Chemistry Chemical Physics}, 21(28):15339--15345, 2019.

\bibitem{Lakhtakia1992}
Akhlesh Lakhtakia.
\newblock {Green's functions and Brewster condition for a halfspace bounded by
  an anisotropic impedance plane}.
\newblock {\em International Journal of Infrared and Millimeter Waves},
  13(2):161--170, feb 1992.

\bibitem{Rotenberg2012}
N.~Rotenberg, M.~Spasenovi{\'{c}}, T.~L. Krijger, B.~le~Feber, F.~J.
  {Garc{\'{i}}a de Abajo}, and L.~Kuipers.
\newblock {Plasmon Scattering from Single Subwavelength Holes}.
\newblock {\em Physical Review Letters}, 108(12):127402, mar 2012.

\bibitem{Picardi2017}
Michela~F. Picardi, Alejandro Manjavacas, Anatoly~V. Zayats, and Francisco~J.
  Rodr{\'{i}}guez-Fortu{\~{n}}o.
\newblock {Unidirectional evanescent-wave coupling from circularly polarized
  electric and magnetic dipoles: An angular spectrum approach}.
\newblock {\em Physical Review B}, 95(24):245416, jun 2017.

\bibitem{Kingsley-Smith2018}
Jack~J. Kingsley-Smith, Michela~F. Picardi, Lei Wei, Anatoly~V. Zayats, and
  Francisco~J. Rodr{\'{i}}guez-Fortu{\~{n}}o.
\newblock {Optical forces from near-field directionalities in planar
  structures}.
\newblock {\em Physical Review B}, 99(23):235410, jun 2019.

\bibitem{Barnett2002}
Stephen~M Barnett.
\newblock {Optical angular-momentum flux*}.
\newblock {\em Journal of Optics B: Quantum and Semiclassical Optics},
  4(2):S7--S16, apr 2002.

\bibitem{Chen2015}
Jun Chen, Jack Ng, Kun Ding, Kin~Hung Fung, Zhifang Lin, and C.~T. Chan.
\newblock {Negative Optical Torque}.
\newblock {\em Scientific Reports}, 4(1):6386, may 2015.

\bibitem{Wang2018}
Pan Wang, Alexey~V. Krasavin, Francesco~N. Viscomi, Ali~M. Adawi,
  Jean-Sebastien~G. Bouillard, Lei Zhang, Diane~J. Roth, Limin Tong, and
  Anatoly~V. Zayats.
\newblock {Metaparticles: Dressing Nano-Objects with a Hyperbolic Coating}.
\newblock {\em Laser \& Photonics Reviews}, 12(11):1800179, nov 2018.

\bibitem{Feng2017}
Tianhua Feng, Yi~Xu, Wei Zhang, and Andrey~E. Miroshnichenko.
\newblock {Ideal Magnetic Dipole Scattering}.
\newblock {\em Physical Review Letters}, 118(17):173901, apr 2017.

\bibitem{Cordova-Castro2019}
R~Margoth C{\'{o}}rdova-Castro, Alexey~V Krasavin, Mazhar~E Nasir, Anatoly~V
  Zayats, and Wayne Dickson.
\newblock {Nanocone-based plasmonic metamaterials}.
\newblock {\em Nanotechnology}, 30(5):055301, feb 2019.

\bibitem{Evlyukhin2011}
Andrey~B. Evlyukhin, Carsten Reinhardt, and Boris~N. Chichkov.
\newblock {Multipole light scattering by nonspherical nanoparticles in the
  discrete dipole approximation}.
\newblock {\em Physical Review B}, 84(23):235429, dec 2011.

\bibitem{Aizpurua2003}
J.~Aizpurua, P.~Hanarp, D.~S. Sutherland, M.~K{\"{a}}ll, Garnett~W. Bryant, and
  F.~J. {Garc{\'{i}}a de Abajo}.
\newblock {Optical Properties of Gold Nanorings}.
\newblock {\em Physical Review Letters}, 90(5):057401, feb 2003.

\bibitem{Mary2007}
A.~Mary, D.~M. Koller, A.~Hohenau, J.~R. Krenn, A.~Bouhelier, and A.~Dereux.
\newblock {Optical absorption of torus-shaped metal nanoparticles in the
  visible range}.
\newblock {\em Physical Review B}, 76(24):245422, dec 2007.

\bibitem{Dutta2008}
Chizuko~M. Dutta, Tamer~A. Ali, Daniel~W. Brandl, Tae-Ho Park, and Peter
  Nordlander.
\newblock {Plasmonic properties of a metallic torus}.
\newblock {\em The Journal of Chemical Physics}, 129(8):084706, aug 2008.

\bibitem{Feng2014}
Hua~Yu Feng, Feng Luo, Renata Kekesi, Daniel Granados, David
  Meneses-Rodr{\'{i}}guez, Jorge~M. Garc{\'{i}}a, Antonio
  Garc{\'{i}}a-Mart{\'{i}}n, Gaspar Armelles, and Alfonso Cebollada.
\newblock {Magnetoplasmonic Nanorings as Novel Architectures with Tunable
  Magneto-optical Activity in Wide Wavelength Ranges}.
\newblock {\em Advanced Optical Materials}, 2(7):612--617, jul 2014.

\bibitem{Born1980}
Max Born and Emil Wolf.
\newblock {\em {Principles of Optics : Electromagnetic Theory of Propagation,
  Interference and Diffraction of Light}}.
\newblock Pergamon Press, Oxford, 6th edition, 1980.

\bibitem{Johnson1972}
P.~B. Johnson and R.~W. Christy.
\newblock {Optical Constants of the Noble Metals}.
\newblock {\em Physical Review B}, 6(12):4370--4379, dec 1972.

\bibitem{Millen2020}
James Millen, Tania~S Monteiro, Robert Pettit, and A~Nick Vamivakas.
\newblock {Optomechanics with levitated particles}.
\newblock {\em Reports on Progress in Physics}, 83(2):026401, feb 2020.

\bibitem{Kuhn2017}
Stefan Kuhn, Alon Kosloff, Benjamin~A. Stickler, Fernando Patolsky, Klaus
  Hornberger, Markus Arndt, and James Millen.
\newblock {Full rotational control of levitated silicon nanorods}.
\newblock {\em Optica}, 4(3):356, mar 2017.

\bibitem{Stickler2018}
Benjamin~A Stickler, Birthe Papendell, Stefan Kuhn, Bj{\"{o}}rn Schrinski,
  James Millen, Markus Arndt, and Klaus Hornberger.
\newblock {Probing macroscopic quantum superpositions with nanorotors}.
\newblock {\em New Journal of Physics}, 20(12):122001, dec 2018.

\bibitem{BEAMS2023}
{BEAMS Software Package, available at the link
  \textit{https://doi.org/10.18742/21975998}}, 2023.

\end{thebibliography}
\bibliographystyle{unsrt}

\newpage
\noindent {\large \textbf{Supplementary Information}}\\

\noindent The development of our package required many programming technicalities that have to be addressed, both to ensure accurate results and to improve performance. The Methods section of the main text discusses the overall process but various technical issues that were considered are detailed in Sections 1-4 of this document. Section 5 defines the polarisation basis vectors used throughout this work. Section 6 provides additional information on how the theoretical force and torque was calculated for the core-shell particle results in the main paper. 

In Sections 1-3, we will discuss the various technical issues that were encountered whilst constructing the package and how they were resolved. We believe this information may prove useful to any reader that wishes to create their own software using our approach. In Section 4, additional information is provided on how the theoretical force and torque was calculated for the core-shell particle results in the main paper. 

\setcounter{section}{0}
\setcounter{equation}{0}
\setcounter{figure}{0}
\renewcommand{\thefigure}{S\arabic{figure}}
\renewcommand{\theequation}{S\arabic{equation}} 
\renewcommand{\thetable}{S\arabic{table}} 

\section{Mesh in k-space}

The angular spectrum approach is fundamentally a summation of plane wave components each with a distinct wavevector, polarisation, amplitude and phase. In the monochromatic case and in the absence of evanescent fields, all possible propagation directions (wavevectors) for a plane wave in 3D space can be mapped onto the surface of a sphere (which we call the k-sphere). 

\begin{figure}[ht]
    \centering
    \includegraphics{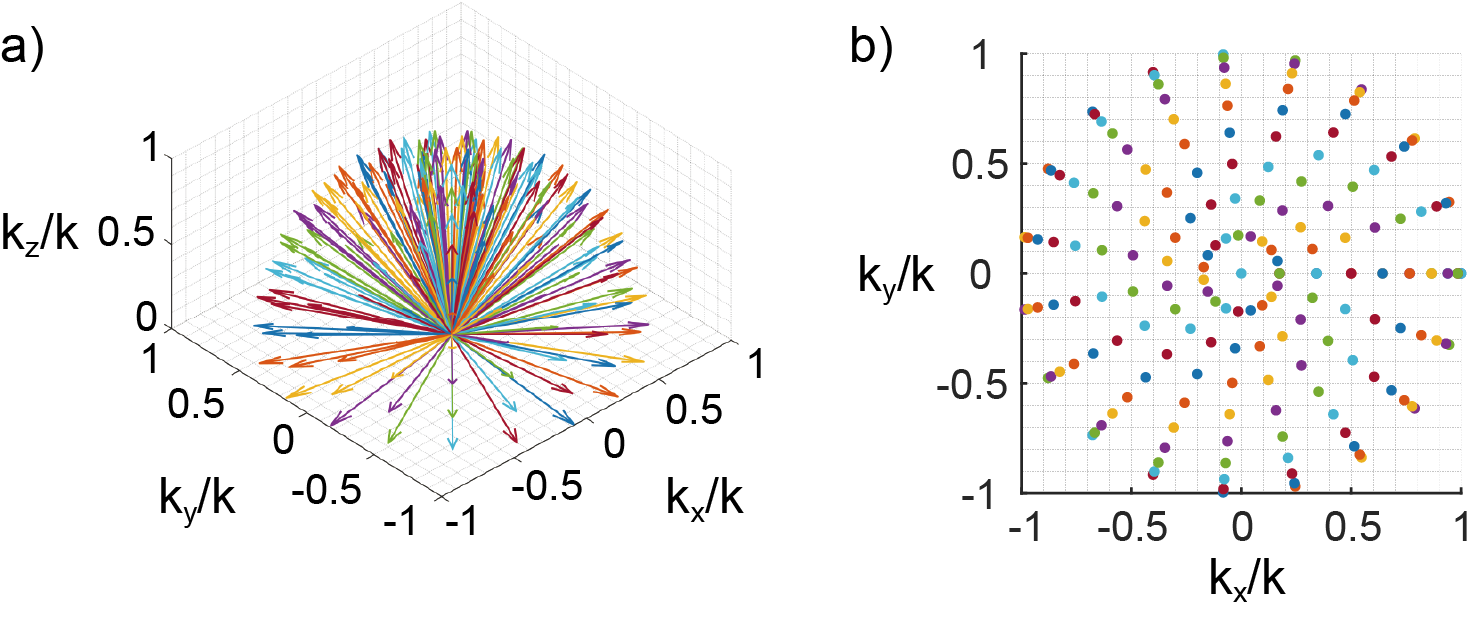}
    \caption{a) A regular sampling of plane wave orientations with a spherical mesh. The arrows trace out the (upper hemisphere) surface of the k-sphere. The arrows have random colours to provide contrast. b) The same spherical mesh is mapped onto the $k_x$-$k_y$ plane.}
    \label{fig:sphericalmesh}
\end{figure}

For the purposes of generating beams incident on axisymmetric structures from plane wave simulations (PWS), one must discretely sample the angular spectrum of the beam. This is done with a regular spherical mesh, depicted in Fig.~\ref{fig:sphericalmesh} because it reduces the likelihood of artefacts arising due to undersampling in a particular direction. The wavevector components $(k_x,k_y,k_z)$ are calculated from spherical mesh coordinates $(\rho,\theta,\phi)$, where $\rho$ is fixed and the angles are defined by,
\begin{align}\label{eq:angledefinitions}
    \theta &=  \text{cos}^{-1}\bigg(\frac{k_z }{k}\bigg), \nonumber \\
    \phi &= \text{tan}^{-1}\bigg(\frac{k_y }{k_x }\bigg).
\end{align}
The number of evenly-spaced samples in $\theta$ and $\phi$ are defined as $N_\theta$ and $N_\phi$ respectively, and are user-defined input parameters. 

Once this sampling mesh is established, optimisations can be made to speed up the calculation time. If the angular spectrum of a beam is highly localised to a small region in the $k_x$-$k_y$ plane, large values for $N_\theta$ and $N_\phi$ may be needed to fully resolve the beam, but this in turn will cause a large number of samples to occur in areas where the angular spectrum has close to zero intensity. Every sample point corresponds to additional computational time but the contributions from these near-zero samples are negligible to the final result and can safely be skipped. The skipping of these negligible sample points is done with the introduction of a cropping parameter $C$ as an optional input for the \textit{MATLAB} package, and takes a value of $0 \leq C < 1$. For an angular spectrum $\mathcal{A}$, a sample point can be denoted $\mathcal{A}_{ij}$. A logical \textit{if}-statement can then check if $\mathcal{A}_{ij}$ should be skipped by computing the inequality,
\begin{equation}
    \frac{\mathcal{A}_{ij}}{\text{max}\{|\mathcal{A}|\}} \geq C.
\end{equation}
If this statement is false, $\mathcal{A}_{ij}$ is deemed to be negligible and is skipped. If it is true,  $\mathcal{A}_{ij}$ is included in the integration of $\mathcal{A}$. In other words, $C = 0.01$ implies that all points in the angular spectrum that have a value less than 1\% of the peak value will be skipped. If $C = 0$, all points in the k-space mesh are included.

\section{Rotation of an angular spectrum}

Starting from a beam with $z$-propagation, the angular spectrum can later be rotated to match the desired beam propagation direction in a process equivalent to rotating the k-sphere. We make use of the well-known matrix $\overset\leftrightarrow{\mathbf{R}}$ for a 3D rotation around a given normalised rotation axis $\hat{\mathbf{u}}=(u_x,u_y,u_z)$ with the general rotation angle $\Theta$, 
\begin{equation}
    \overset\leftrightarrow{\mathbf{R}} = 
    \begin{pmatrix}
        \text{cos}(\Theta) + u_x^2g(\Theta) 
        & u_x u_y g(\Theta) -u_z \text{sin}(\Theta) 
        & u_x u_z g(\Theta) +u_y \text{sin}(\Theta) \\
        u_y u_x g(\Theta) +u_z \text{sin}(\Theta)  
        & \text{cos}(\Theta) + u_y^2g(\Theta) 
        & u_y u_z g(\Theta) -u_x \text{sin}(\Theta) \\
        u_z u_x g(\Theta) -u_y \text{sin}(\Theta) 
        & u_z u_y g(\Theta) +u_x \text{sin}(\Theta) 
        & \text{cos}(\Theta) + u_z^2g(\Theta)  \\
    \end{pmatrix}
\end{equation}
where $g(\Theta) = \big(1-\text{cos}(\Theta)\big)$
and the identity for rotating a generic vector field, 
\begin{equation}\label{eq:vectorfieldrotation}
    \mathbf{V}(\mathbf{r}) = \overset\leftrightarrow{\mathbf{R}} \, \mathbf{V}^{\hat{\mathbf{K}} \parallel \hat{\mathbf{z}}} \bigg (\overset\leftrightarrow{\mathbf{R}}^{-1}\mathbf{r}\bigg),
\end{equation}
where $\mathbf{V}^{\hat{\mathbf{K}} \parallel \hat{\mathbf{z}}}$ is a vector field propagating along the $z$-axis described by the real-space coordinates $\mathbf{r}$, and $\mathbf{V}$ is the desired vector field. 

When calculating the fields on a beam incident on a scattering structure, the angular spectrum $\mathcal{A}$ is numerically sampled into a discrete set of points. Each point in $\mathcal{A}$ corresponds to a PWS with a particular incident $\mathbf{k}$. Therefore, each point must be mapped to the $\mathbf{k}$ of the incident plane wave in the relevant simulation data $\mathbf{E}_{\text{PWS}}$. This mapping is provided by two vector field rotations which preserve the simulation's polarisation state. These rotations are simplified by using the convention that all numerical PWS raw data have their plane wave incident in the $xz$-plane, and that any material structure in the simulation must have its axis of continuous rotational symmetry around the $z$-axis. The rotation axes are therefore the $y$-axis for the $\theta$ rotation (when the structure is spherical) and then the axisymmetric $z$-axis for the $\phi$ rotation. If the structure lacks spherical symmetry, the plane wave must be simulated for all values of $\theta$ and then all the results are rotated around the $z$-axis to match the required values of $\phi$ and complete the integration in k-space. 

\section{Angular spectra of beam classes}

Since our approach of obtaining the angular spectrum of a beam always begins with the beam propagating along $z$, it is convenient to pre-calculate the angular spectrum of some 
types of beams commonly used in optics. These are shown in Tab.~\ref{tab:beams}. The polarisation of a beam with a homogeneous polarisation in the transverse plane is denoted by $\text{\boldmath$\eta_\perp$}=\eta_x \, \hat{\mathbf{x}} + \eta_y \, \hat{\mathbf{y}}$. The beam waist is $w$ and the functions $H_l(x)$ and $L_m^l(x)$ correspond to Hermite and Laguerre polynomials respectively, with argument $x$ and orders $l$ and $m$. These expressions are derived 
\begin{table}[h]
    \centering
    \caption{The transverse electric field distribution in the focal plane for various types of paraxial beams and their corresponding angular spectra, obtained from Eq.~(5). The beam axis is denoted by $\mathbf{K}$, and sgn$(l)$ is the sign function of $l$ (i.e. sgn$(l)=+1$ if $l>1$ and sgn$(l)=-1$ if $l<1$. The expression for the Laguerre-Gaussian angular spectrum is not valid if $l<0$ and $m>0$.}
    \resizebox{\textwidth}{!}{%
    \begin{tabular}{|c|c|c|}
        \hline
        Beam type & $\mathbf{E}_\perp^{\hat{\mathbf{K}} \parallel \hat{\mathbf{z}}}$ & $\mathcal{F}_{\perp}^{\hat{\mathbf{K}} \parallel \hat{\mathbf{z}}}$ \\
        \hline
        Gaussian & $\text{\boldmath$\eta_\perp$} e^{-\frac{x^2+y^2}{w^2}}$ & $ \text{\boldmath$\eta_\perp$} \frac{w^2}{4 \pi} e^{-\frac{w^2(k_x^2+k_y^2)}{4}} $ \\
        Hermite-Gaussian & $ \text{\boldmath$\eta_\perp$} H_l\Big(\frac{\sqrt{2} x}{w}\Big) H_m\Big(\frac{\sqrt{2}y}{w}\Big) \, e^{-\frac{x^2+y^2}{w^2}} $ & $ \text{\boldmath$\eta_\perp$} \frac{w^2}{4 \pi} H_l\Big(\frac{k_x w}{\sqrt{2}}\Big) H_m\Big(\frac{k_y w}{\sqrt{2}}\Big)  \, e^{-\frac{i \pi }{2}(l+m)} e^{-\frac{w^2(k_x^2+k_y^2)}{4}} $ \\
        Laguerre-Gaussian & $ \text{\boldmath$\eta_\perp$} \Big(\frac{\sqrt{x^2+y^2}}{w}\Big)^{|l|} L_m^l\Big(\frac{2(x^2+y^2)}{w^2}\Big) e^{il \, \text{tan}^{-1}(\frac{y}{x})} e^{-\frac{x^2+y^2}{w^2}} $ & $\text{\boldmath$\eta_\perp$} \frac{w^2}{4 \pi} L_m^l\Big(\frac{w^2(k_x^2+k_y^2)}{2}\Big) (-1)^m \big(-\frac{i w}{2}(k_x + \text{sgn}(l) \, i k_y) \big)^l  \, e^{-\frac{w^2(k_x^2+k_y^2)}{4}}$ \\
        Azimuthal & $\frac{2\sqrt{2}}{w} e^{-\frac{x^2+y^2}{w^2}} (y \hat{\mathbf{x}} -x \hat{\mathbf{y}}) $ & $-\frac{i w^3}{2\sqrt{2} \pi} e^{-\frac{w^2(k_x^2+k_y^2)}{4}} (k_y \hat{\mathbf{x}} -k_x \hat{\mathbf{y}})$ \\
        Radial & $\frac{2\sqrt{2}}{w} e^{-\frac{x^2+y^2}{w^2}} (x \hat{\mathbf{x}} +y \hat{\mathbf{y}}) $ & $-\frac{i w^3}{2\sqrt{2} \pi} e^{-\frac{w^2(k_x^2+k_y^2)}{4}} (k_x \hat{\mathbf{x}} + k_y \hat{\mathbf{y}})$ \\
        \hline
    \end{tabular} 
    }
    \label{tab:beams}
\end{table}

Note that all these angular spectra $\mathcal{F}_{\perp}^{\hat{\mathbf{K}} \parallel \hat{\mathbf{z}}}$ can be later converted into the piecewise angular spectra $\mathcal{A}$ which is used in Eq.~(1) from the main text to calculate non-paraxial fields. The longitudinal field can be reconstructed for all of these beams using Eq.~(7) from the main text.

\section{Polarisation basis decomposition}

Eq.~(2) of the main text shows the angular spectrum of a beam being decomposed into plane waves of two orthogonal polarisation states. This step makes use of the orthogonal $p$ and $s$ polarisation basis unit vectors $\hat{\mathbf{e}}_p$ and $\hat{\mathbf{e}}_s$ defined as,
\begin{align}\label{eq:ethephvectors}
    \hat{\mathbf{e}}_p &= 
    \begin{pmatrix}
        \text{cos}(\theta) \, \text{cos}(\phi)\\
        \text{cos}(\theta) \, \text{sin}(\phi)\\
        \text{sin}(\theta)
    \end{pmatrix} \equiv
    \frac{1}{k \, k_t}
    \begin{pmatrix}
        k_x \, k_z\\
        k_y \, k_z\\
        -k_t^2
    \end{pmatrix}, & 
    \hat{\mathbf{e}}_s &= 
    \begin{pmatrix}
        -\text{sin}(\phi)\\
        \text{cos}(\phi)\\
        0
    \end{pmatrix} \equiv
    \frac{1}{k_t}
    \begin{pmatrix}
        -k_y\\
        k_x\\
        0
    \end{pmatrix}.
\end{align}
These are the usual polar and azimuthal unit vectors in spherical coordinates. 
In our approach, these vectors are used to decompose the 3D vector field $\mathcal{A}$ into the complex scalar fields $A_p$ and $A_s$ that appear in Eq.~(2) of the main text. This is computed via dot products, $A_p = \mathcal{A} \cdot \hat{\mathbf{e}}_p$ and $A_s = \mathcal{A} \cdot \hat{\mathbf{e}}_s $. Note that $\mathcal{A}$ must be a full 3D angular spectrum with longitudinal components included. If the longitudinal component is missing, $A_{p,s}$ will be inaccurate for tightly focused beams.

\section{Mesh in real-space}

In a numerical simulation (e.g. PWS), the electromagnetic field is usually sampled by a discrete real-space cubic mesh. Our method requires the rotation of this field and ultimately the sum of many rotated versions of this field. 
These rotations are not so straightforward for a discrete vector field because if one were to rotate a cubic mesh by a small angle, most of the mesh points would not align with the original points. This means that the discrete fields cannot be summed since there will be a mismatch in the $xyz$ coordinates. This summation is necessary in order to perform the integration of the angular spectrum in Eq.~(3) of the main text. In essence, all the rotated electromagnetic field data must be defined on the same coordinate mesh. 

Our solution is to rotate the original real-space mesh $\mathbf{r}_{\text{PWS}}$ in the opposite direction to which the vector field will be rotated,
\begin{equation}
    \mathbf{r}^\text{R} =  \overset\leftrightarrow{\mathbf{R}}_\phi(-\phi) \, \overset\leftrightarrow{\mathbf{R}}_\theta(-\theta) \, \mathbf{r}_{\text{PWS}}.
\end{equation}
It is important to note that the axes of rotations and the order of the rotations must be chosen to preserve the polarisation state of any rotated fields (i.e. the polar angle unit vectors $\hat{\mathbf{e}}_p$ and azimuthal angle unit vectors $\hat{\mathbf{e}}_s$ must be preserved). 
For a package ingesting PWS data with the plane wave incident along $z$, first the $\theta$ rotation occurs around the negative $y$-axis and after that, the $\phi$ rotation occurs around the $z$-axis. 
The original unrotated vector field $\mathbf{E}_{\text{PWS}}$ is then interpolated onto this backwards-rotated mesh $\mathbf{r}^\text{R}$, such that $\mathbf{E}_{\text{PWS}}(\mathbf{r}) \to \mathbf{E}_{\text{PWS}}(\mathbf{r}^\text{R})$. So long as the resolution of the mesh is sufficiently subwavelength, linear regression was found to provide adequate interpolation results. The interpolated fields and the new mesh are then forward-rotated to the desired orientation. This ensures that the final mesh is identical to the original PWS mesh, and all the various rotated fields can be integrated. 

These real-space rotations also give rise to another issue. Since the simulated fields are evaluated on a simulation box with a given size, some data points near the corners of the mesh may rotate into a position that is outside of the boundaries of the original mesh. This can cause the interpolation algorithm to inaccurately extrapolate out to this region and introduce a large error in these points. To compensate for this, the mesh for the final integrated fields must be smaller than the mesh of the simulation, such that any rotation of the smaller mesh is contained within the simulation mesh. 
The user provides the coordinates that they wish to use for the optical beam (the beam mesh) and the package uses the data from the simulations (defined on the simulation's mesh) to generate the beam. The two meshes are not related in any way, and linear interpolation acts as the bridge between the two. 

Since the rotations always occur around the centre of the mesh, the furthest a point on a cube can get from the rotation axis is when the point is located on a vertex. If the beam's mesh is a cube with side length $d_{\text{beam}}$ and the simulation's mesh is a cube with side length $d_{\text{PWS}}$, the following inequality must be satisfied,
\begin{equation}\label{eq:simulationsizeinequality}
    d_{\text{PWS}} \geq \sqrt{3} \,  d_{\text{beam}}.
\end{equation}
This corresponds to a spherical locus traced out by the vertices of the cube under all possible 3D rotations. If this relation is not met, the vertices of the beam's field distribution are prone to large inaccuracies.

\section{Theoretical force and torque on the core-shell particle}

Fig.~1d in the main text demonstrates some optical force and torque results that can be obtained with our beam-generation approach. Since the particle in question is a subwavelength body with spherical symmetry, Mie theory is employed to provide an alternative calculation method with which to compare and judge reliability of the results. Mie theory is a technique widely used in scattering problems and describes the scattering behaviour of a sphere via a series of electric and magnetic Mie coefficients, $a_n$ and $b_n$ respectively [43,44]. When the particle's size is subwavelength as is the case of Fig.~1d, the first-order dipole coefficients $a_1$ and $b_1$ are dominant. The electric and magnetic dipole polarisabilities $\alpha_e$ and $ \alpha_m$ respectively are related to the Mie coefficients by,
\begin{align}\label{eq:miecoefficients}
        \alpha_e &= a_1 \frac{i 6 \pi \varepsilon}{k^3}, &
        \alpha_m &= b_1 \frac{i 6 \pi}{k^3}.
\end{align}
From the polarisabilities, electric and magnetic dipole moments $\mathbf{p}$ and $\mathbf{m}$ respectively can be calculated using,
\begin{align}\label{eq:dipolesfromalpha}
    \mathbf{p} &= \alpha_e  \mathbf{E}, &
    \mathbf{m} &= \alpha_m  \mathbf{H}.
\end{align}
In this section, $\mathbf{E}$ and $\mathbf{H}$ are strictly the fields incident on the particle, as opposed to the total fields including the particle's scattering. 

The following equations can then be used to calculate the theoretical force and torque, $\mathbf{F}_\text{th}$ and $\mathbf{N}_\text{th}$ respectively, on the spherical particle considered in Fig.~1 of the main text,
\begin{equation}\label{eq:dipoleforce}
    \langle \mathbf{F}_\text{th} \rangle = \frac{1}{2} \mathfrak{Re} \bigg\{ (\mathbf{\nabla} \otimes  \mathbf{E}) \mathbf{p}^* + \mu (\mathbf{\nabla} \otimes  \mathbf{H}) \mathbf{m}^* - \frac{k^4}{6 \pi \varepsilon c} (\mathbf{p} \times \mathbf{m}^*) \bigg\},
\end{equation}
\begin{equation}\label{eq:theoreticaltorque}
    \langle \mathbf{N}_\text{th} \rangle = \frac{1}{2}\mathfrak{Re} \bigg\{ \big(\mathbf{p}^* \times \mathbf{E} + \mu  \mathbf{m}^* \times \mathbf{H}\big) - \frac{k^3}{6 \pi} \bigg(\frac{1}{\varepsilon } \mathfrak{Im}\{\mathbf{p}^* \times \mathbf{p}\} + \mu  \mathfrak{Im}\{\mathbf{m}^* \times \mathbf{m}\} \bigg) \bigg\}.
\end{equation}
In this case, the only non-zero components of $\mathbf{F}_\text{th}$ and $\mathbf{N}_\text{th}$ are the $\hat{\mathbf{z}}$-components with magnitudes $F_\text{th}=0.959$ nN/W and $N_\text{th}=44.0$ aNm/W (once normalised by the beam power).

\end{document}